\documentclass[a4paper,UKenglish]{lipics-v2016}
\usepackage{microtype}%if unwanted, comment out or use option "draft"

\newcommand{\nolineover}[2]{\genfrac{}{}{0pt}{}{#1}{#2}}

\newtheorem{Definition}{Definition}
\newtheorem{Lemma}{Lemma}
\newtheorem{Theorem}{Theorem}
\newtheorem{Note}{Note}
\newtheorem{Observation}{Observation}
\newtheorem*{Proof}{Proof}

\title{A linear lower bound for incrementing a space-optimal integer 
representation in the bit-probe model
\footnote{
       The author acknowledges support from the Danish National Research Foundation
   and The National Science Foundation of China (under the grant 61361136003) for
       the Sino-Danish Center for the Theory of Interactive Computation.
}}
\titlerunning{A linear lower bound for incrementing a space-optimal integer representation}

\author[1]{Mikhail Raskin }
\affil[1]{
 {Department of Computer Science, Aarhus University}
     \\ 
     \texttt{raskin@mccme.ru} 
}

\authorrunning{M. Raskin}

\Copyright{Mikhail Raskin}

\subjclass{F.2.2 Nonnumerical Algorithms and Problems, G.2.1 Combinatorics}

\keywords{
 binary counter, data structure, integer representation,
 bit-probe model, lower bound.
}

\begin{document}
\maketitle

\begin{abstract}

We present the first linear lower bound for the number of bits required
to be accessed in the worst case to increment an integer in an arbitrary
space-optimal binary representation.  The best previously known lower bound
was logarithmic.  It is known that a logarithmic number of read bits in the
worst case is enough to increment some of the integer representations that
use one bit of redundancy, therefore we show an exponential gap between
space-optimal and redundant counters.

Our proof is based on considering the increment procedure for a space optimal
counter as a permutation and calculating its parity. For every space optimal
counter, the permutation must be odd, and implementing an odd permutation
requires reading at least half the bits in the worst case.  The combination
of these two observations explains why the worst-case space-optimal problem
is substantially different from both average-case approach with constant
expected number of reads and almost space optimal representations with
logarithmic number of reads in the worst case.

\end{abstract}

\section{Introduction}

We consider the problem of representing integers in a certain
range by binary codes to support efficient increment operations 
in the bit-probe model. Our main interest is representing 
integers in the range $0,\ldots, 2^n-1$ by exactly $n$ bits.

Most computational tasks require storing integers, 
and in some cases
special encodings are better suited to the task. 
Probably the first encodings used in bit-based memory are
the standard positional binary notation and binary coded decimal.
We will consider the task of incrementing an integer counter in the
bit probe model. Following [Fredman1978], 
an increment algorithm is defined 
as a \textbf{decision assignment tree} (DAT), a binary tree 
where each inner node specifies a bit of the code 
that has to be read to make a decision, and every leaf node
contains a set of changes to the code to perform.
In this model the number 
of bits read  by an increment
is the depth of the corresponding leaf,
and the number of bits written is the number
of changes in the leaf.

The standard binary notation with $n$ bits
reads and writes only 2 bits on average,
but has to read and write $n$ bits in the worst case.
It is also \textbf{space-optimal}, i.e. every 
combination of bits represents a unique integer.
Gray codes [Gray1953] allow writing only one 
bit for each increment operation,
but they still require reading all $n$ bits.
In the article [Fredman1978] Fredman introduces 
the notion of DAT and
considers codes that can use more bits than
necessary but still require writing only a single bit per update.
A logarithmic lower bound is proven for the number of bits read
in the worst case for such a code. This bound is sometimes 
cited in connection with the codes that write a constant
number of bits per increment; while the bound is correct,
the proof in [Fredman1978] does use the fact that only one
bit is written. Fredman also gave a construction of a code
such that the increment procedure needs to read only 
$O(\log n)$ bits in the worst case, but the code length 
is worse than for the standard notation by a large multiplicative
factor.
Frandsen, Miltersen and Skyum consider a more general problem 
of encoding elements of a generic monoid; their article [FMS1997]
provides a general lower bound for the number of bits read by an
increment procedure for integers
in the worst case and a construction of a code
with an increment procedure that needs to read $\log_2 n+1$ bits 
in the worst case. This code needs $\log_2 n$ extra bits.
Constructions developed by Bose et al. in [BCJMMS2010] and 
Rahman and Munro [RM2010] require a single extra bit or less 
and achieve logarithmic number of bits read by the corresponding 
increment procedures. 
Rahman and Munro also prove a lower bound of $\Omega(\sqrt{n})$
bits read in the worst case for the special case when increment
reads a different subset of bits for every input or at least 
each subset of bits is read only for a constant number of 
inputs (this is a strong condition; for example, the standard 
binary notation reads only the last bit for half of all the inputs).
Elmasry and Katajainen  provide
a code [EK2013] 
that uses a logarithmic number of extra bits and requires
the increment procedure to read 
only a logarithmic number of bits in the worst case
while ensuring
efficient implementation on a word-based RAM machine.

For the space-optimal case 
there is a code [BGPS2014] that allows reading 
$n-1$ bits in the worst case
and writes no more than $3$ bits for each increment operation.
This code has been found by brute force search.
The best previously known lower bound on the number of 
bits to read in the worst case given in [FMS1997] 
is logarithmic in~$n$.

\begin{table}
\centering
\begin{tabular}{p{8em}|p{12em}|p{12em}}
Number of~bits read to~increment
        & Space-optimal & Single extra bit
\\
        \hline
Average &   
        \multicolumn{2}{|c|}{ $\Theta(1)$ (binary notation)}
        \\
\hline
Worst-case &
    \parbox[t]{12.5em}{
            \mbox{$\geqslant \log_2 n$} [FMS1997]  \\
            \mbox{$\leqslant n-1$} [BGPS2014]
            \vskip2mm
            \mbox{$\geqslant \frac{n}{2}$} \textbf{our contribution}
            \\ ~
    }
    & 
        \parbox[t]{12.5em}{
        \mbox{$\geqslant\log_2 n$ } [FMS1997] \\
        \mbox{$\leqslant\log_2 n+O(1)$} [RM2010]
       }
\end{tabular}
\caption{A summary of best known results}
\label{results}
\end{table}

In this paper we close the exponential gap and 
settle the complexity of the problem
up to a multiplicative factor of 2 by proving
that every representation of integers from $0$ to $2^{n}-1$
using $n$ bits require
the increment operation to read at least 
$ \frac{n}{2} $ bits in the 
worst case. The proof uses properties of 
permutations to explain why 
the space-optimal codes and the redundant codes 
are qualitatively different 
from the point of view of the worst-case complexity 
of the increment operation.

In the next section we give the standard definitions related 
to permutations and cite their standard properties. Then we 
define our model of computation. In the section 3 we give 
an overview of the core ideas and an outline of the proof.

The section 4 contains the detailed definitions of the 
constructions and the proofs of their basic properties.
The section 5 contains the combinatorial details
and the final part of the proof.
The purpose of the sections 4 and 5 is to remove any remaining 
uncertainty after the brief presentation of the proof in the 
section 3.

\section{Preliminaries}

\subsection{Algebraic preliminaries}

In this subsection we recall algebraic
notions and the standard theorems about permutations
required by our proof.
We follow the definitions from [Lauritzen2003],
but equivalent
definitions can be found in many other abstract algebra books.
This subsection can be skipped at the reader's discretion.

\begin{Definition}
A \textbf{group} is a pair $(G,\circ)$ consisting of 
a set $G$ and a function $\circ: G\times G\to G$, such that
the following conditions hold:
\\ \phantom{\quad} Associativity: 
$\forall a,b,c\in G: a\circ(b\circ c) = (a\circ b)\circ c$
\\ \phantom{\quad} 
Neutral element:
$\exists e\in G: \forall a\in G: e\circ a = a\circ e=a$\\
~ \phantom{\qquad }
$e$ is called a \textbf{neutral element} of a group
\\ \phantom{\quad} 
Inverse element:
$\forall a\in G: \exists b\in G: a\circ b=b\circ a=e$ \\
~ \phantom{\qquad}
$b$ is called an \textbf{inverse element} of $a$
\end{Definition}

\begin{Theorem}
There can be only one neutral element in a group $G$. 
Let $e$ denote the unique neutral element in the group.
Also, for every $a\in G$ there can be only one inverse element.
Let $a^{-1}$ denote the unique inverse element of $a$.
The inverse element of a composition 
$a\circ b$ is $b^{-1}\circ a^{-1}$.
\end{Theorem}

\begin{Definition}
The \textbf{symmetric group} $S_n$ is 
the group of all bijections (one-to-one correspondences)
from the set $\{1,2,3,\ldots,n\}$ to itself. 
The group operation $\circ$
is the composition of functions. 
Each bijection in the symmetric
group is called a \textbf{permutation}. 
The neutral element in the symmetric group is 
the \textbf{identity permutation}
$\sigma(x)=x$. The inverse element of a permutation~$\sigma$
is the inverse function $\sigma^{-1}$, also called 
the \textbf{inverse permutation} of $\sigma$.
%Two permutations $\sigma, \tau\in S_n$ are called 
%\textbf{disjoint}
%if $\forall x\in\{1,\ldots,n\}: (\sigma(x)=x) \vee (\tau(x)=x)$,
%i.e. every element is left in its place by at least one of the 
%permutations.
A permutation $\sigma\in S_n$
can be 
denoted~by~$
\left(\nolineover{1}{\sigma(1)}\nolineover{2}{\sigma(2)}\nolineover{3}{\sigma(4)}
\cdots\nolineover{n}{\sigma(n)}\right)$.
In the present paper $\circ$ will always be written explicitly.
\end{Definition}

\begin{Definition}
Suppose we are given 
$k\leqslant n$
different elements $x_1, \ldots, x_k\in \{1,\ldots,n\}$.
A permutation $\sigma$ given by $$
\sigma(x_1)=x_2, \quad
\sigma(x_2)=x_3, \quad 
\sigma(x_3)=x_4, \quad 
\ldots, \quad
\sigma(x_{k-1})=x_k, \quad 
\sigma(x_k)=x_1 
$$ and $\sigma(x)=x$ if $x\notin\{x_1,\ldots,x_k\}$ is
called a $k$-\textbf{cycle}. 
It is denoted $\sigma=(x_1 \, x_2 \, \ldots \, x_k)$.
A $2$-cycle is also called a \textbf{transposition}.
\end{Definition}

%\textbf{Theorem}
%Every permutation can be represented as a composition
%of disjoint cycles.

\begin{Definition}
Let $\sigma\in S_n$ be a permutation. A pair of indices $(i,j)$
where $1\leqslant i < j\leqslant n$ is called an \textbf{inversion}
(of the permutation $\sigma$) if $\sigma(i)>\sigma(j)$ (i.e.
if the permutation inverts the order in which $i$ and $j$ go).
\end{Definition}

\begin{Definition}
A permutation is called \textbf{even} if it has an even number 
of inversions;
otherwise it is called \textbf{odd}.
\end{Definition}

\begin{Definition}
The composition of two even permutations or two odd permutations
is an even permutation. The composition of an even permutation
and an odd permutation in any order is an odd permutation.
A $k$-cycle is an odd permutation if $k$ is even and 
an even permutation if $k$ is odd.
The inverse of a permutation has the same parity.
\end{Definition}

\subsection{The model}

\begin{Definition}
A \textbf{space-optimal code} is an
encoding function ${\mathit{Enc}}$
from $\{0,1,2,\ldots,2^n-1\}$
to the set of bit sequences of length $n$. 
Each code implicitly defines
the decoding function ${\mathit{Dec}}={\mathit{Enc}}^{-1}$ and the 
\textbf{increment function}
${\mathit{Inc}}(x)={\mathit{Enc}}({\mathit{Dec}}(x)+1)$.
We will also call such codes \textbf{counters}.
\end{Definition}

\begin{Definition} \label{def-DAT}
A \textbf{decision assignment tree} (DAT) is a binary tree 
where each inner node specifies a single position in the code 
and every leaf node contains a set of changes (assignments)
in the code.
Execution of a DAT on an input bit sequence starts in the root 
node and then the next node is the left child of the current
node if the bit in the specified position 
of the input code is $0$
and the right child of the current node otherwise. 
When a leaf node is reached, the output is calculated by 
taking the input and
setting the bits in the positions specified for this 
leaf node to the specified values.
\end{Definition}

Each DAT defines a function from bit sequences of some length
to bit sequences of the same length. We will say that
all the nodes visited during execution of DAT on some input
(including the root and the
leaf node) \textbf{handle} this input.
The number of bits read is the depth of the corresponding leaf,
and the number of bits written to the code is the number
of assignments in the leaf.

\begin{Definition}
The set $\{0,1\}^n$ is called an $n$-dimensional
\textbf{hypercube}. An element of a hypercube
is called a \textbf{vertex}. 
Every vertex has $n$ coordinates. 
A $k$-dimensional \textbf{face} is a subset of 
the $n$-dimensional hypercube
defined by specifying the values of some $n-k$ coordinates
(we will call these coordinates \textbf{fixed})
and allowing all the possible combinations of values
of the remaining $k$ coordinates (we will call 
these coordinates \textbf{free}).
Each vertex of a hypercube
is a bit sequence; we will identify each vertex with the
integer it represents in the standard binary notation.
The order on the hypercube vertices 
given by comparing the 
vertices as integers is called the 
\textbf{lexicographic order}.
\end{Definition}

\section{The bound and the proof outline}

The main result of the present paper is:
the increment function for every space-optimal code
representing integers from $0$ to $2^{n}-1$ must read
at least $\frac{n}{2}$ bits in the worst
case. In other words, there is no space optimal code
such that the corresponding increment function never
reads more than $L(n):=\frac{n}{2}-1$ bits.

In this section we present an informal outline of the proof.
The core idea of the proof is representing the permutation
specified by $\mathit{Inc}$ as a composition of two permutations, 
$\mathit{Before}$ and $\mathit{After}$, defined in terms 
of vertices
 handled by the same leaf node 
in the DAT implementing
$\mathit{Inc}$.

Assume that for some $n$ there is a way to encode 
integers such that the
corresponding increment function ${\mathit{Inc}}$ can 
be implemented by a DAT that reads at most $L(n)$ bits 
in the worst case.
Without loss of generality we can consider an implementation
that always reads exactly $L(n)$ bits.
The increment function maps
the $n$-dimensional hypercube into itself and can be
considered as a permutation. 
This permutation is a cycle of length $2^n$, and, therefore,
an odd permutation.

We will use a representation of the increment function
as a composition of two permutations, $\mathit{Before}$
and $\mathit{After}$. 
Each leaf of the DAT implementing $\mathit{Inc}$ handles
some $(n-L(n))$-dimensional 
face of the $n$-dimensional hypercube.
By definition, the restriction of $\mathit{Inc}$ on each of these
faces changes some of the bits in the same way for all the
vertices in the face. We also know that $\mathit{Inc}$ 
is a bijection, therefore only the fixed bits can be changed.
This can be interpreted as a parallel translation
of the face.
The image of each of the
faces in $F$ under $\mathit{Inc}$ is again 
a $(n-L(n))$-dimensional face.
Let $F$ denote the set of all the faces handled by any 
leaf of
the DAT. The set of their images, $\mathit{Inc}(F)$, is also
a set of faces. 
Every vertex lies in exactly one face
from $F$ and in exactly one face from $\mathit{Inc}(F)$.
Let's fix some order of enumeration of $F$,
i.e. $F=\{F_0, F_1, \ldots, F_{2^{n-L(n)}-1}\}$. This 
also defines an order on $\mathit{Inc}(F)$, namely,
$\mathit{Inc}(F) = \{\mathit{Inc}(F_0),\ldots,\mathit{Inc}(F_{2^{n-L(n)}-1})\}$.
We can consider three orders on the hypercube: the 
standard lexicographic ordering; the ordering 
where we compare two vertices by first compare
their corresponding faces in $F$ and fall back to 
lexicographic order inside each face; and the ordering
where we first compare the containing faces from $\mathit{Inc}(F)$.
We can enumerate all the vertices of the hypercube 
according to these three orders. Enumeration in the
lexicographic order is the identity permutations.
The remaining two orders give non-trivial permutations;
we will call them $\mathit{Before}$ and $\mathit{After}$.
Note that 
$\mathit{Inc}=\mathit{After}\circ \mathit{Before}^{-1}$, 
because the $i$-th vertex
in the $j$-th face of $F$ is by definition of $F$ mapped
to the $i$-th vertex of the $j$-th face of $\mathit{Inc}(F)$.

We prove that 
$\mathit{Before}$ and $\mathit{After}$ are both even.
The proof is the same for both permutations.
We need to calculate the number of inversions. Every 
face is enumerated in lexicographic order, so there 
can be no inversion including two vertices from the same face.
When we consider two different faces, they do not intersect and 
therefore have to have a coordinate
which is fixed in both the faces and has a different value.
There have to be at least two common free coordinates between 
the faces if each of them has $L(n)<\frac{n}{2}$ fixed bits and
at least one fixed bit position is shared.
Faces with two common free coordinates have an even
number of inversions using one vertex from each of the faces, because
inversions where the less significant common free coordinate doesn't 
affect the comparison come in multiples of four (two bits in the 
less significant common position can be flipped at will), and
those where the bits in the less significant common position matter 
come in the multiples of two (the coordinates in the more significant
common position must match, and it doesn't matter if both are equal to
zero or to one).
Therefore the total number
of inversions in the permutation $\mathit{Before}$ 
(the same holds for
the permutation $\mathit{After}$) is even.

But if $\mathit{Before}$ and $\mathit{After}$ are both even,
$\mathit{Inc}=\mathit{After}\circ \mathit{Before}^{-1}$ has to be even.
The contradiction proves that our initial assumption
was wrong and the implementation of ${\mathit{Inc}}$
has to read at least $\frac{n}{2}$ bits
in the worst case.

\section{Defining the constructions used in the proof}

\begin{figure}[h]
\includegraphics{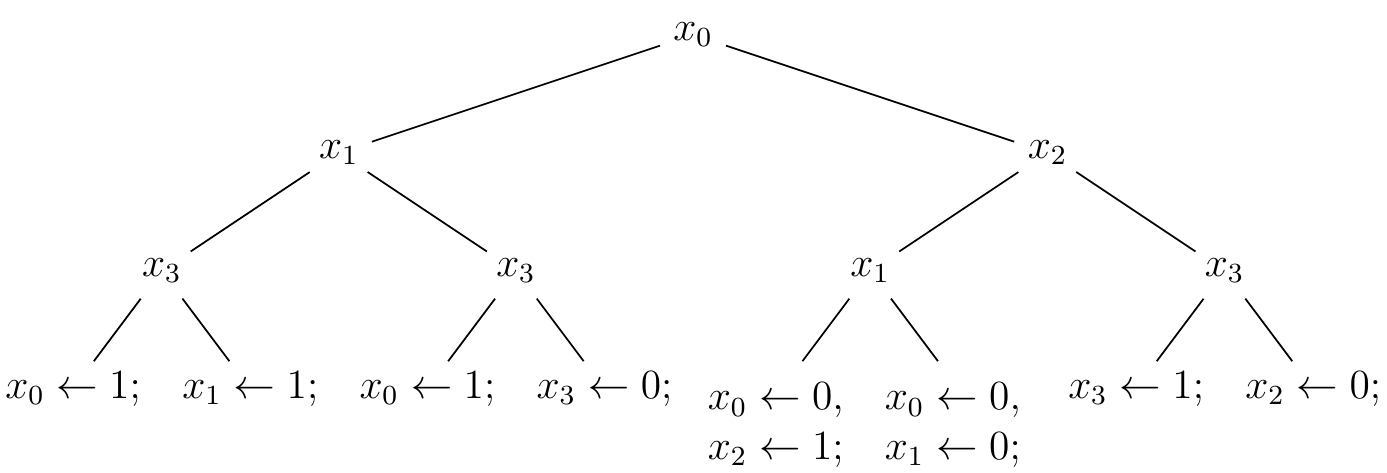}
\caption{The decision tree from [BGPS2014]}
        \label{dat-bpgs2014}
\end{figure}

To illustrate some of the notation, we will use the space-optimal
integer representation from [BGPS2014]. This representation
was initially found by a brute-force search.
Its increment function was presented as a decision 
assignment tree (figure \ref{dat-bpgs2014}).

\subsection{The cycle defined by the increment function}

\begin{Observation} 
        If the increment function can be 
described by a decision assignment tree (DAT), it can be 
represented by a DAT of the same depth with all the leaves
having the same depth.
\end{Observation} 

\begin{Lemma} 
        For a space-optimal representation 
of integers in the range $\{0,\ldots,{2^n-1}\}$
the increment
function is a bijection of the set of bit strings of length $n$.
By interpreting these strings as integers
using standard binary notation,
we can represent the increment function
as a permutation of $\{0,1,2,\ldots,2^n-1\}$. 
This permutation is a cycle of length $2^n$.
This cycle is an odd permutation.
\end{Lemma} 

\begin{Proof} 
The increment function $\mathit{Inc}$ maps $\mathit{Enc}(2^n-1)$ to $\mathit{Enc}(0)$, 
$\mathit{Enc}(0)$ to $\mathit{Enc}(1)$, $\mathit{Enc}(1)$ to $\mathit{Enc}(2)$, etc.
$\mathit{Enc}(0), \ldots, \mathit{Enc}(2^n-1)$ 
are all the different binary strings of length $n$
with each string used exactly once, 
so $\mathit{Inc}$ is a bijection.
If we interpret the bit strings as integers we get
the cycle $(\, \mathit{Enc}(0) \, \mathit{Enc}(1) \, \ldots \, \mathit{Enc}(2^n-1)\,)$.
The cycle is an odd permutation because its length is even.

\begin{figure}[h]
        \center
        \includegraphics[width=10cm]{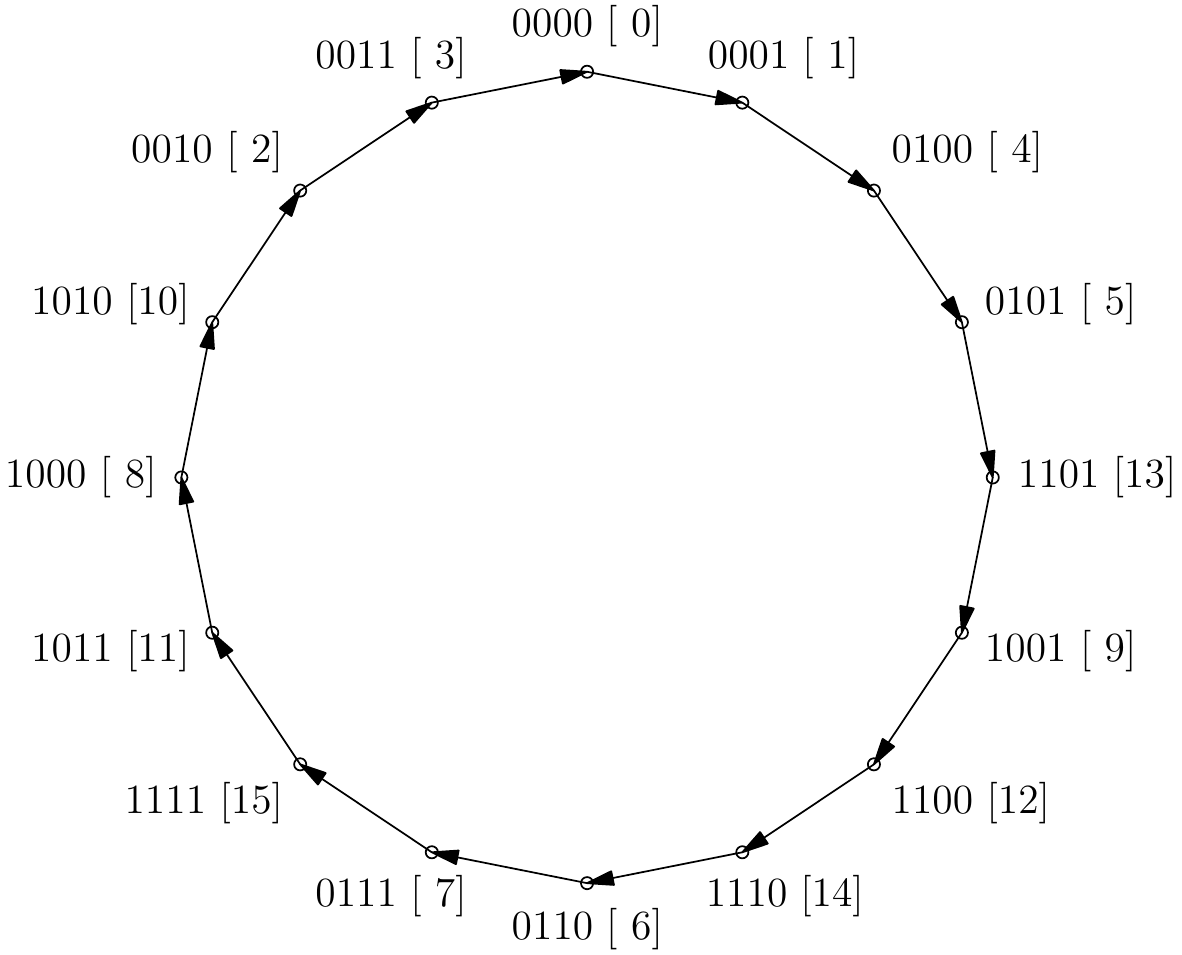}
\caption{The cycle corresponding to the example from [BGPS2014]}
\end{figure}
 
In the example from [BGPS2014] the cycle is as shown 
in the figure.
The «first» ($x_0$) bit from the algorithm's
explanation is used as the least significant bit.
 We can also write this cycle as 
 a table of $\mathit{Inc}$ function values:\\
 \begin{tabular}{|c|c|c|c|c|c|c|c|}
 \hline 
  code &  \textit{Inc}(code)   &   code   &  \textit{Inc}(code)   
& code &  \textit{Inc}(code)   &   code   &  \textit{Inc}(code)   \\
 \hline
 0000 [ 0] & 0001 [ 1] & 0100 [ 4] & 0101 [ 5] & 1000 [ 8] & 1010 [10] & 1100 [12] & 1110 [14] \\
 0001 [ 1] & 0100 [ 4] & 0101 [ 5] & 1101 [13] & 1001 [ 9] & 1100 [12] & 1101 [13] & 1001 [ 9] \\
 0010 [ 2] & 0011 [ 3] & 0110 [ 6] & 0111 [ 7] & 1010 [10] & 0010 [ 2] & 1110 [14] & 0110 [ 6] \\
 0011 [ 3] & 0000 [ 0] & 0111 [ 7] & 1111 [15] & 1011 [11] & 1000 [ 8] & 1111 [15] & 1011 [11] \\
 \hline
 \end{tabular}

~

The standard notation for this permutation is
 $ \left( 
 \nolineover{ 0}{1}
 \nolineover{ 1}{4}
 \nolineover{ 2}{3}
 \nolineover{ 3}{0}
 \nolineover{ 4}{5}
 \nolineover{ 5}{13}
 \nolineover{ 6}{7}
 \nolineover{ 7}{15}
 \nolineover{ 8}{10}
 \nolineover{ 9}{12}
 \nolineover{10}{2}
 \nolineover{11}{8}
 \nolineover{12}{14}
 \nolineover{13}{9}
 \nolineover{14}{6}
 \nolineover{15}{11}
 \right) $, the cycle notation is 
$(0\,1\,4\,5\,13\,9\,12\,14\,6\,7\,15\,11\,8\,10\,2\,3)$.
This permutation has 39 inversions, so it is
an odd permutation.
\end{Proof}

\subsection{Decision assignment trees and the corresponding faces}

\begin{Lemma} 
        Given a DAT for a $n$-bit integer
representation and a node of depth $k$, the set of 
all inputs handled (in the sense of definition \ref{def-DAT}) by the chosen node is 
an~$(n-k)$-dimensional
face.
\end{Lemma} 

\begin{Proof} 
The proof goes by induction. 
The root has depth $0$ and handles 
all the hypercube, which can be considered an $n$-dimensional 
face. A child of node of depth $k$ has depth $k+1$; the set 
of vertices handled by the child node can be obtained from 
the set of vertices handled by the parent node by fixing 
the coordinate inspected in the parent node 
to one of the two possible values. This coordinate was 
a free coordinate, so we get an $(n-k-1)$-dimensional
face out of a $(n-k)$-dimensional one.
\end{Proof} 

\begin{Lemma}
        If a DAT implements a bijection, every coordinate
in every assignment in the leaf nodes
is a fixed coordinate of the
face handled by the corresponding node,
i.e. this coordinate is inspected
in one of the ancestor nodes of the leaf node.
\end{Lemma}

\begin{Proof} 
All the vertices need to have different images.
Assume a leaf node handled two vertices 
which differ in the assigned coordinate.
This leaf node would handle some face
containing both vertices, and this face
would also contain some 
two vertices that differ 
\textit{only} in the assigned 
coordinate. But these two latter vertices 
would have the same image, and this is not allowed.
\end{Proof} 

\begin{Lemma}
        If a DAT implements a bijection, 
the image of the face
handled by a leaf node 
is a translation of this face. In particular, the
image is also a face of the hypercube.
\end{Lemma}

\begin{Proof} 
        Changing some of the fixed coordinates of a face 
performs a parallel translation.

We will now illustrate how the faces are moved.
The $4$-bit counter using $3$ reads for 
every increment corresponds 
to a 4-dimensional hypercube. It is more convenient to draw it as
two 3-dimensional cubes side by side (so the extra coordinate is
projected to the vector proportionate to the projection of the 
first coordinate). In this case the faces corresponding to the
decision assignment tree (DAT) leaves are 1-dimensional faces.
We will represent the faces corresponding to the DAT leaves by
solid lines.
Both pictures use the same set of arrows to represent the movement
of the faces. The top picture shows the faces before the moves, and
the bottom picture shows the faces after the moves.
 
 \begin{figure}[h]
 \includegraphics{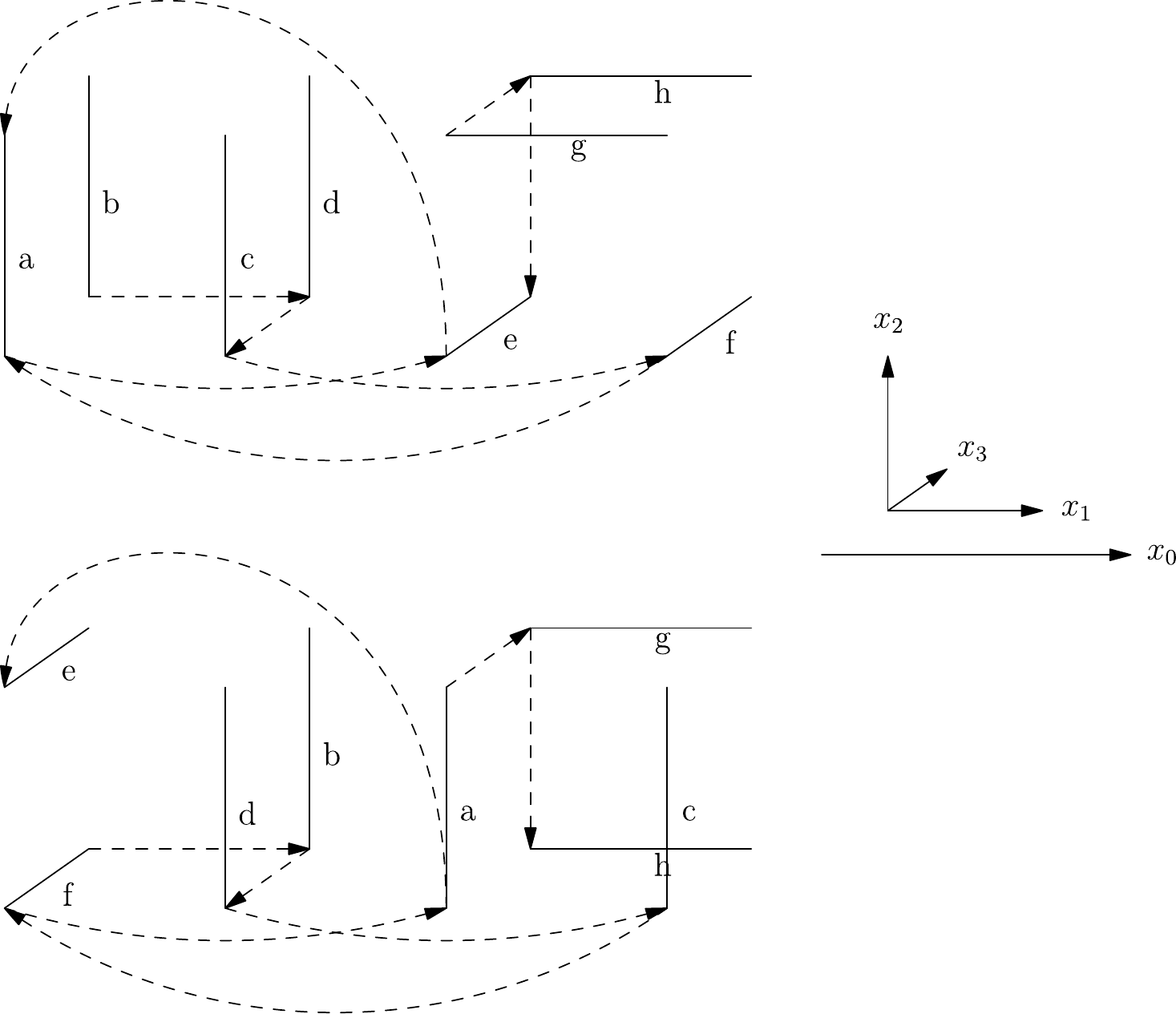}
         \caption {The translations corresponding to the counter.}
\end{figure}
 
We can list the vertices by face. Initially they are split 
in the following way: \\
 \begin{tabular}{|c|c|c|c|c|c|}
 \hline
 face & vertices & vertices (decimal) &  face & vertices & vertices (decimal) \\
 \hline
 a  &  0000 and 0100 & 0 and 4  & e  &  0001 and 1001 & 1 and 9 \\
 b  &  1000 and 1100 & 8 and 12  & f  &  0011 and 1011 & 3 and 11  \\
 c  &  0010 and 0110 & 2 and 6  & g  &  0101 and 0111 & 5 and 7 \\
 d  &  1010 and 1110 & 10 and 14  & h  &  1101 and 1111 & 13 and 15 \\
 \hline
 \end{tabular}
  
 ~ 
 
 and after the faces are moved we get a new split:\\
 \begin{tabular}{|c|c|c|c|c|c|}
 \hline
 face & vertices & vertices (decimal) &  face & vertices & vertices (decimal) \\
 \hline
 a  & 0001 and 0101  &   1 and  5   & e  & 0100 and 1100  &  4 and 12    \\
 b  & 1010 and 1110  &  10 and 14   & f  & 0000 and 1000  &  0 and  8    \\
 c  & 0011 and 0111  &   3 and  7   & g  & 1101 and 1111  & 13 and 15   \\
 d  & 0010 and 0110  &   2 and  6   & h  & 1001 and 1011  &  9 and 11    \\
 \hline
 \end{tabular}
\end{Proof} 

\subsection{Enumerating the faces and the vertices}

Assume we have a balanced DAT implementing the increment function $\mathit{Inc}$
for a space-optimal integer representation. 
Each leaf handles the
vertices forming a face, these faces are disjoint, 
have the same dimension
and cover the entire hypercube. The $\mathit{Inc}$-images of these faces are again 
disjoint faces of the same dimension covering the entire hypercube.
We need to choose some order on the faces 
handled by different leaves; 
it is not important which order we use so we will use the order of leaves
in the DAT.

\begin{Definition} 
Let $F_i$ denote the $i$-th face in the chosen order.
\end{Definition} 

\begin{Definition} 
Let $\mathit{Inc}$ be a permutation implemented by a balanced DAT
of depth $l$.
The $\mathit{Before}$ permutation is the enumeration of all the
vertices in the hypercube by first enumerating all the vertices in $F_0$ in the
lexicographic order,
then all the vertices in $F_1$, etc. In general, 
the $j$-th vertex in the lexicographic order
on the face $F_i$ will have 
the number $i\times2^l+j$.
We could write $\mathit{Before(i\times2^l+j)}=F_i[j]$.
The $\mathit{After}$ permutation is defined in a similar way with
the $j$-th vertex in the face $\mathit{Inc}(F_j)$ having the 
number $i\times2^l+j$. 
\end{Definition} 

\begin{Lemma}
The $\mathit{Inc}$ permutation is the composition of permutations
$\mathit{Before}^{-1}$ and $\mathit{After}$, i.e.
$\mathit{Inc}=\mathit{After}\circ\mathit{Before}^{-1}$. This can also
be written as $\forall k \in \{0,\ldots,2^n-1\}:
\mathit{Inc}(\mathit{Before}(k))=\mathit{After}(k)$.
\end{Lemma}

\begin{Proof}
Let $k$ be represented as $i\times2^l+j$.
We have $\mathit{Inc}(\mathit{Before}(k))=\mathit{Inc}(\mathit{Before}(i\times2^l+j))=
\mathit{Inc}(F_i[j])=\mathit{Inc}(F_i)[j]=\mathit{After}(i\times2^l+j)=\mathit{After}(k)$
(the $j$-th element in the $i$-th face gets translated together with the entire face).

 For the example algorithm the $\mathit{Before}(\cdot)$ numbering is:
 $$\left(
 \nolineover{0000}{0000}
 \nolineover{0001}{0100}
 \nolineover{0010}{1000}
 \nolineover{0011}{1100}
 \nolineover{0100}{0010}
 \nolineover{0101}{0110}
 \nolineover{0110}{1010}
 \nolineover{0111}{1110}
 \nolineover{1000}{0001}
 \nolineover{1001}{1001}
 \nolineover{1010}{0011}
 \nolineover{1011}{1011}
 \nolineover{1100}{0101}
 \nolineover{1101}{0111}
 \nolineover{1110}{1101}
 \nolineover{1111}{1111}
 \right)$$
 in binary, or 
 $\left(
 \nolineover{ 0}{0}
 \nolineover{ 1}{4}
 \nolineover{ 2}{8}
 \nolineover{ 3}{12}
 \nolineover{ 4}{2}
 \nolineover{ 5}{6}
 \nolineover{ 6}{10}
 \nolineover{ 7}{14}
 \nolineover{ 8}{1}
 \nolineover{ 9}{9}
 \nolineover{10}{3}
 \nolineover{11}{11}
 \nolineover{12}{5}
 \nolineover{13}{7}
 \nolineover{14}{13}
 \nolineover{15}{15}
 \right)$ 
 in decimal notation.
 The $\mathit{After}$ numbering is
 $$\left(
 \nolineover{0000}{0001}
 \nolineover{0001}{0101}
 \nolineover{0010}{1010}
 \nolineover{0011}{1110}
 \nolineover{0100}{0011}
 \nolineover{0101}{0111}
 \nolineover{0110}{0010}
 \nolineover{0111}{0110}
 \nolineover{1000}{0100}
 \nolineover{1001}{1100}
 \nolineover{1010}{0000}
 \nolineover{1011}{1000}
 \nolineover{1100}{1101}
 \nolineover{1101}{1111}
 \nolineover{1110}{1001}
 \nolineover{1111}{1011}
 \right)$$
 or $\left(
 \nolineover{ 0}{1}
 \nolineover{ 1}{5}
 \nolineover{ 2}{10}
 \nolineover{ 3}{14}
 \nolineover{ 4}{3}
 \nolineover{ 5}{7}
 \nolineover{ 6}{2}
 \nolineover{ 7}{6}
 \nolineover{ 8}{4}
 \nolineover{ 9}{12}
 \nolineover{10}{0}
 \nolineover{11}{8}
 \nolineover{12}{13}
 \nolineover{13}{15}
 \nolineover{14}{9}
 \nolineover{15}{11}
 \right)$.
 We first enumerate the two vertices in the $a$ face, then 
 the two vertices in the $b$ face, etc. using the positions 
 of the faces before and after the move, respectively.
 We can see that the $\mathit{Before}$ permutation is odd and 
 the $\mathit{After}$ permutation is even. As the example algorithm
 reads more than a half of all the bits, getting an odd permutation
is possible.
 
 As an illustration, 10 is the first vertex in the $d$ face before the
translation of the face by the increment function, 
$\mathit{Inc}(10)=2$. The first vertex in the face $d$ (the fourth face)
has number $6$ in the $\mathit{Before}$ ordering; we see that
$\mathit{Before}^{-1}(10)=6$. After the shift the first vertex in the face $d$
is 2. We see that $\mathit{After}(6)=2$ and $\mathit{Inc}(10) = 
\mathit{After}(\mathit{Before}^{-1}(10)) =
(\mathit{After} \circ \mathit{Before}^{-1})(10) = 2$.
\end{Proof}
 
\section{Calculating the parity of the permutations}

Both of the permutations 
$\mathit{Before}$ and $\mathit{After}$ are specified in the same way, 
by cutting the hypercube
into faces and enumerating the faces. 
It is now sufficient to show that any permutation
specified in that way is even if the faces
have no more than $L(n)=\frac{n}{2}-1$ 
fixed coordinates.
We will prove that the $\mathit{After}$ permutation is even;
exactly the same proof will work for the $\mathit{Before}$
permutation.

\subsection{Faces and the inversions}

Recall that an inversion of a permutation $\sigma$ is a pair 
of numbers $x<y$ such that $\sigma(x)>\sigma(y)$.

\begin{Lemma} \label{no-inversions-same}
There are no inversions of the permutation $\mathit{After}$
such that $\mathit{After}(x)$ and $\mathit{After}(y)$ are in the
same face.
\end{Lemma}

\begin{Proof} 
If $\mathit{After}(x)$ and $\mathit{After}(y)$ are in the
same face and $x<y$ then $x$ has a lower number inside the
face than $y$. But the face is enumerated in the
lexicographic order,
so $\mathit{After}(x)<\mathit{After}(y)$.
\end{Proof}

\begin{Lemma} \label{even-inversions-cross}
Consider two faces, $\mathit{Inc}(F_i)$ and 
$\mathit{Inc}(F_{i'})$ such that $i<i'$. 
The number of inversions such that 
$\mathit{After}(x)\in\mathit{Inc}(F_{i})$ 
and
$\mathit{After}(y)\in\mathit{Inc}(F_{i'})$ 
is even.
\end{Lemma}

\begin{Note}
If $i>i'$ then there are no inversions because $x$ would 
always be larger than~$y$.
\end{Note}

\begin{Proof}
Note that the vertices are enumerated face-by-face, so
the condition that 
$\mathit{After}(x)\in\mathit{Inc}(F_{i})$ 
and
$\mathit{After}(y)\in\mathit{Inc}(F_{i'})$ 
guarantees $x<y$. Every vertex in each face has exactly one
number, so we can just count the number of pairs $(u, v)$ where
$u\in \mathit{Inc}(F_{i})$
and 
$v\in \mathit{Inc}(F_{i'})$
and 
$u>v$.
We will use the assumption that each of the faces has
$L(n)=\frac{n}{2}-1$
fixed coordinates. This means that no more than $n-2$ coordinates
are fixed in any of the two faces. Therefore there are at least 
two coordinates that are free for both faces.
We can write the faces' coordinates one on top of the other one:
 $
 \nolineover{0}{0}\nolineover{0}{*}\nolineover{*}{1}\cdots
 $
Let us consider the least significant of the common free coordinates 
and call it $p$. 

We will split all the pairs $(u,v)$ into two groups based on the number of 
coordinates that have to be checked 
to perform a lexicographic comparison 
if we read from the most significant bit. Either it is enough to 
read only some of the coordinates that are 
more significant than $p$, or we need to 
to read the coordinate $p$ and maybe some more.
\\1) The number of pairs of codes where 
the comparison can be made without
 considering the bits on the position~$p$
 (and less significant positions) is
 even, because half of these codes have $0$ 
 in the first face on the
 position $p$ and the other half have $1$.
\\2)If we have to consider the bits 
in the position $p$, the bits in every more 
significant common position must be equal. There 
is an even number of such pairs because changing 
a pair of bits in the same position from $0,0$ to $1,1$ doesn't affect the 
comparison.
 
We have split all the pairs with $u>v$ into two even-sized sets,
as we have to consider either only bits more significant that the 
position $p$ or the bits including position $p$. 
Therefore the total number of such pairs is even.
 
This finishes the proof that the number of inversions containing two
vertices in the two given faces is even.
\end{Proof}

\subsection{Summarizing the inversion counts}

\begin{Lemma}
The $\mathit{After}$ permutation (and the $\mathit{Before}$ permutation)
for a $\mathit{Inc}$ function represented by a DAT of depth less than
$\frac{n}{2}$ are even.
\end{Lemma}

\begin{Proof}
In the previous subsection we have proven that every inversion of the
$\mathit{After}$ permutation has to include elements from different
faces (lemma \ref{no-inversions-same}).
We have also proven that for every pair of faces the number of
inversions represented by their elements is even 
(lemma \ref{even-inversions-cross}). If we sum the inversions 
for all the pairs of faces we get all the inversions of the permutation.
Therefore the number of inversions of the permutation is even.
\end{Proof}

Now we can prove the main theorem.

\begin{Theorem}
The increment function for every space-optimal binary code
representing integers from $0$ to $2^{n}-1$ must read
at least $\frac{n}{2}$ bits in the worst
case. In other words, there is no space optimal binary code
such that the corresponding increment function never
reads more than $L(n):=\frac{n}{2}-1$ bits.
\end{Theorem}

\begin{Proof}
If there is an increment function for a space-optimal
binary integer
representation reading at most
$L(n)$ bits in the worst case, the corresponding 
$\mathit{Before}$ and $\mathit{After}$ permutation would 
both be even. Then the $\mathit{Inc}$ function would be 
an even permutation. But the increment function for a 
space-optimal binary integer representation has to 
be a cycle of length $2^n$, i.e. an odd permutation.
The  contradiction proves that our assumption was impossible.
\end{Proof}

\section{Handling a weaker definition of increment}

Our definition of increment assumed that incrementing the largest value always yields
zero. This requirement can be removed from the definition.

\begin{Theorem}
Consider an increment 
procedure for a space-optimal integer representation that correctly handles all the 
possible values except the maximum and always leaves at least two bits unread.
Such a procedure always maps the encoding of the maximum value to the encoding of zero.
\end{Theorem}

\begin{Proof}
Let $n$ denote the total amount of bits in the code.
Let us assume that the increment procedure applied to the maximum values
doesn't yield zero.
Let $k$ denote a position where the encoding of zero and the result of 
incrementing the encoding of the maximum value differ. Without loss of generality
we can assume that the encoding of zero has $0$ at the position $k$ and
the value $a=Inc(Enc(2^n-1))$ has $1$ at the position $k$.

Let us count the vertices $x$ such that $Inc(x)$ has the value $1$ at the position $k$.
Almost all vertices have exactly one preimage, $Enc(0)$ has no 
preimages and $a$ has $2$ preimages, so the answer should be $2^{n-1}+1$.
On the other hand, a face corresponding to a vertex of the DAT can have no, all or half 
of its vertices in the set, depending on the orientation; in any case this is an even
number and the total sum has to be an even number.

The contradiction proves that our assumption is false and incrementing
the maximum values has to yield zero as the result.
\end{Proof}

\section{Future directions}

Minor tweaks of the presented proof allow to extend 
the result to cover
nondeterministic increment procedures. For $n>10$ the same
bound can be proven even if we allow arbitrary changes 
of the inspected bits together with an arbitrary 
reversible linear transformation of the unread bits
in every leaf node of a DAT.

Closing the gap between the $\frac{n}{2}$ lower bound and $n-1$ 
upper bound remains an open problem. Our conjecture is that the 
true value is $n-o(n)$.

\section{Acknowledgements}

I am grateful to Gerth Brodal for attracting attention to this problem
and his help with editing the present paper.
I am extremely grateful to Gudmund Frandsen for a lot of useful discussions 
and the efforts he has spent on reading multiple draft versions of this proof.
I am grateful to the anonymous reviewers of this and previous versions of
the present paper for their valuable advice regarding presentation.
I am grateful to an anonymous reviewer for the suggestion that the existence of
common fixed coordinates for disjunct faces improves the bound by one; and 
for the suggestion that the image of the maximum element can be proven to be 
zero even if this assumption is not included in the definition.

\section{References}

\noindent
\textbf{[BGPS2014]}
Gerth Stølting Brodal, Mark Greve, Vineet Pandey, Srinivasa Rao Satti.
Integer representations towards efficient counting in the bit probe model. \textit{J. Discrete Algorithms 26: 34-44 (2014)}

\noindent
\textbf{[Gray1953]} 
Frank Gray. Pulse code communications. U.S. Patent (2632058), 1953.

\noindent
\textbf{[Fredman1978]} 
Michael L. Fredman. 
Observations on the Complexity of Generating Quasi-Gray Codes.
\textit{SIAM J. Comput., 7(2), 134–146}

\noindent
\textbf{[FMS1997]}
Gudmund Skovbjerg Frandsen, Peter Bro Miltersen and Sven Skyum.
Dynamic word problems. 
\textit{J. ACM 44, 257–271 (1997)}

\noindent
\textbf{[BCJMMS2010]} 
Prosenjit Bose, Paz Carmi, Dana Jansens, Anil Maheshwari,
Pat Morin, and Michiel Smid.
Improved Methods For Generating Quasi-gray Codes.
\textit{Proceedings of 12th Scandinavian Symposium and Workshops on Algorithm Theory 
(SWAT 2010), 224–235}

\noindent
\textbf{[RM2010]} M. Ziaur Rahman, J. Ian Munro.
Integer Representation and Counting in the Bit Probe Model.
\textit{Algorithmica (2010) 56: 105–127}

\noindent
\textbf{[EK2013]}
Amr Elmasry, Jyrki Katajainen. In-Place Binary Counters
\textit{Proceeding of 38th International Symposium on 
Mathematical Foundations of Computer Science (MFCS 2013), 349–360}

\noindent
\textbf{[Lauritzen2003]} 
Niels Lauritzen. Concrete Abstract Algebra. 
\textit{First edition: 2003}

\end{document}